\documentclass[aps,prl,twocolumn,notitlepage,superscriptaddress]{revtex4-1}
\usepackage[T1]{fontenc}
\usepackage{amsmath}
\usepackage{amssymb}
\usepackage{lipsum}

\usepackage[unicode=true,colorlinks=true,citecolor=blue,urlcolor=blue]{hyperref}

\usepackage{bm}
\usepackage{color}
\usepackage{epsfig}

\usepackage[normalem]{ulem}

\renewcommand {\phi}{{\varphi}}

\begin{document}
\title{Dynamical stabilization by vacuum fluctuations in a cavity: \\ Resonant electron scattering in the ultrastrong light-matter coupling regime}

\author{D.~A. Zezyulin}
\affiliation{Department of Physics, ITMO University, St.~Petersburg, 197101, Russia}

\author{S.~A. Kolodny}
\affiliation{Department of Physics, ITMO University, St.~Petersburg, 197101, Russia}
\affiliation {Department of Applied and Theoretical Physics, Novosibirsk~State~Technical~University,
Karl~Marx~Avenue~20,~Novosibirsk~630073,~Russia}
\author{O.~V. Kibis}
\affiliation {Department of Applied and Theoretical Physics, Novosibirsk~State~Technical~University,
Karl~Marx~Avenue~20,~Novosibirsk~630073,~Russia}
\author{I.~V. Tokatly}
\affiliation{Nano-Bio
  Spectroscopy group and European Theoretical Spectroscopy Facility (ETSF), Departamento de Pol{\'i}meros y Materiales Avanzados: F\'isica, Qu\'imica y Tecnolog\'ia, Universidad del
  Pa\'is Vasco, Av. Tolosa 72, E-20018 San Sebasti\'an, Spain}
 \affiliation{IKERBASQUE, Basque Foundation for Science, 48009 Bilbao, Spain}
\affiliation{Donostia International Physics Center (DIPC), 20018 Donostia-San Sebasti\'{a}n, Spain}
\affiliation{Department of Physics, ITMO University, St.~Petersburg, 197101, Russia}

\author{I.~V. Iorsh}
\email{i.iorsh(c)metalab.ifmo.ru}
\affiliation{Department of Physics, ITMO University, St.~Petersburg, 197101, Russia}
\affiliation {Department of Applied and Theoretical Physics, Novosibirsk~State~Technical~University,
Karl~Marx~Avenue~20,~Novosibirsk~630073,~Russia}

\begin{abstract}
We developed a theory of electron scattering by a short-range repulsive potential in a cavity.  In the regime of ultrastrong electron coupling to the cavity electromagnetic field, the vaccuum fluctuations of the field result in the dynamical stabilization of a quasi-stationary polariton state confined in the core of the repulsive potential. When the energy of a free electron coincides with the energy of the confined state, the extremely efficient resonant non-elastic scattering of the electron accompanied by emission of a cavity photon appears. This effect is discussed as a basis for possible free-electron sources of non-classical light.
\end{abstract}

\maketitle
{\it Introduction.} Engineering the properties of quantum systems by dynamical modulation of their parameters based in the Floquet theory (the Floquet engineering) became an established research field with a plethora of  fundamental predicted and observed~ phenomena~\cite{1,2,3,4,5,6}. In many setups, the role of periodic time-dependent modulation is played by the intense off-resonant electromagnetic field which does not excite any transitions in the system but merely renormalizes its parameters in a controllable fashion. This effect termed the electromagnetic dressing appears to be especially powerful in application to nanostructures, where various electromagnetically induced phase transitions --- metal-insulator transitions~\cite{7,8,9}, topological phase transitions~\cite{10,Iorsh,11,12}, etc --- were considered.

Among many dressing-field effects, the dynamical stabilization of electronic states in nanostructures by an electromagnetic field~\cite{13} should be noted especially. The essence of the effect is that the dressing of a static repulsive potential by a high-frequency electromagnetic field renormalizes it in such a way that the potential acquires a local minimum in its core. Since the repulsive potential dressed by the field turns into attractive one near its center, it can confine quasi-stationary electron states stabilized by the field. Such a dynamical stabilization leads to many peculiar effects, including the electromagnetically induced electron pairing~\cite{13,14}, the features of electron transport~\cite{14_1,14_2}, etc. Despite having direct analogs in classical mechanics (e.g., the famous Kapitza pendulum~\cite{15}), the effect of electromagnetically induced dynamical stabilization is relatively new. This may be attributed to its deeply non-perturbative nature inaccessible for the standard perturbation analysis. The non-perturbative character of the effect also requires relatively high field intensities which are rather undesirable for possible experimental demonstrations since they would lead to inevitable heating. The question arises, if a similar effect can be realized without an intense external illumination and solely due to the coupling of electron to vacuum fluctuations of the electromagnetic field in a properly engineered photonic cavity.
 \begin{figure}[!t]
    \centering
    \includegraphics[width=1.0\columnwidth]{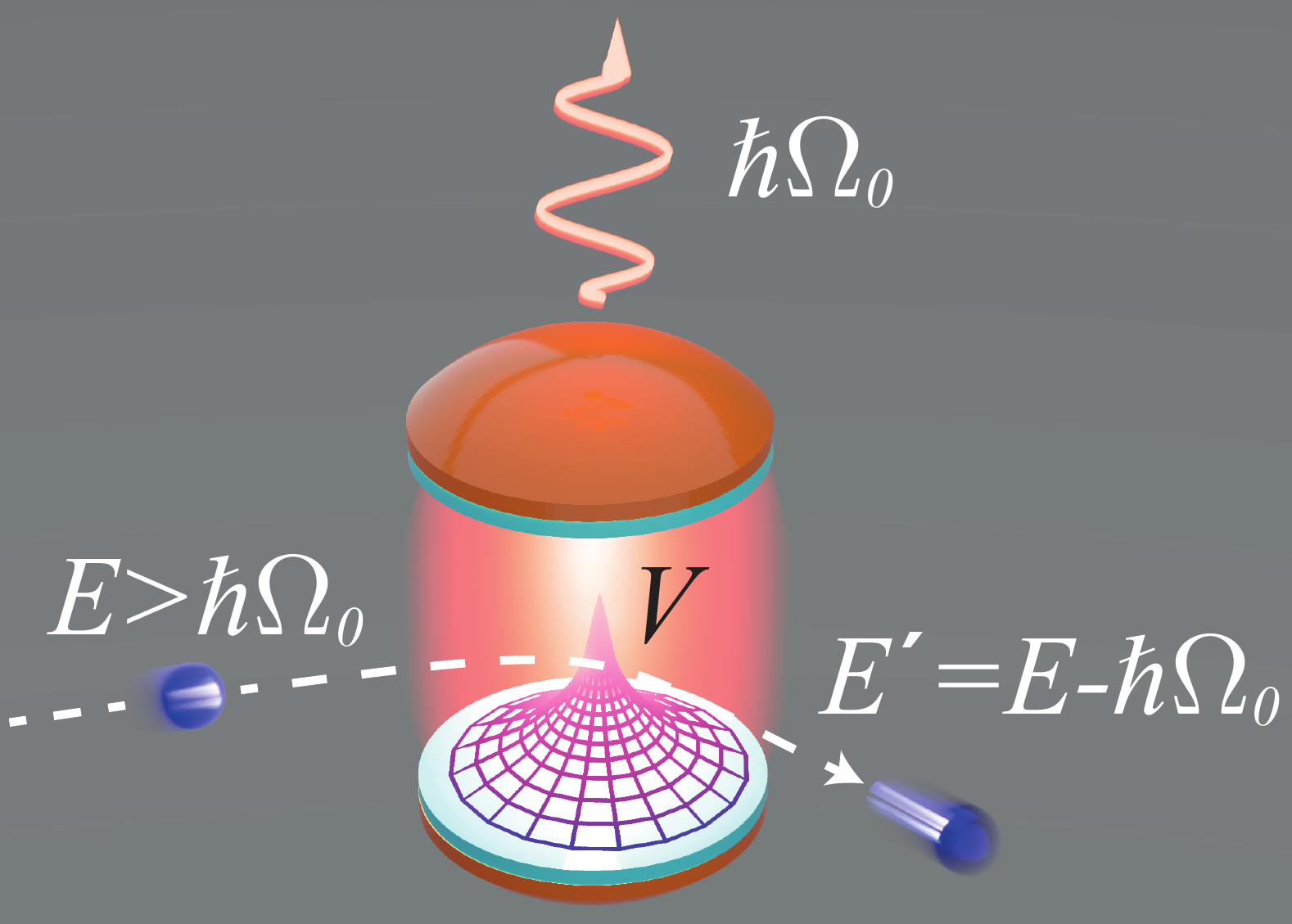}
    \caption{Sketch of the system under consideration: A free electron with the energy $E$ enters a single mode cavity with the resonant frequency $\Omega_0$ and scatters on a short-range repulsive potential $V$, emitting the cavity photon.
    }
    \label{fig:geometry}
\end{figure}

The research on the interaction of matter with a quantized electromagnectic field in optical cavities --- the cavity quantum electrodynamics (CQED) --- has recently enjoyed the surge of interest caused by the qualitative improvements in the nanofabrication technology. These improvements allowed to push the characteristic energies of light-matter interaction in nanosystems embedded into cavities to the values comparable to the cavity photon energy~\cite{16,17}. In the regime of ultrastrong light-matter interaction~\cite{18}, the coupling of a material system to vacuum fluctuations of electromagnetic field modifies the ground state of the system. This peculiar phenomena and its consequences led to the emergence of the new research field known as a cavity-QED materials engineering~\cite{19,20,21,21_a,21_b}. There were predictions of numerous fascinating cavity-induced phase transitions, including superconductivity ~\cite{22,23,24,25,26}, ferroelectric phase transitions~\cite{27}, topological phase transitions~\cite{28,29} as well as substantial modification of the chemical reactions inside the cavity~\cite{30,31,32,33,34,35,36}. In the present research, we show that the ultrastrong light-matter coupling regime can facilitate the dynamical stabilization of a polariton state confined at a repulsive potential. As a result, the resonant scattering of an electron through the state --- which is accompanied by the efficient photon emission --- appears.

{\it Model.} We consider an electron with the energy $E$, which enters a single mode cavity with the resonant frequency $\Omega_0$ and scatters on a short-range repulsive potential $V$, emitting the cavity photon of the energy $\hbar\Omega_0$ (see Fig.~\ref{fig:geometry}). The short range scattering potential for free electron can be realized with a charged metallic of dielectric nanoparticle (e.g., a nanosphere or a nanorod).

For simplicity, we will restrict the following analysis by the case of the one-dimensional scattering potential $V(x)=V_0\delta(x)$, where $V_0>0$. Then the Hamiltonian of the considered system reads
\begin{align}\label{H0}
  \hat{\cal H}_0=\frac{1}{2m}(-i\hbar \partial_x-e \hat{A})^2+V_0\delta(x)+\hbar\Omega_0(\hat{a}^{\dagger}\hat{a})
\end{align}
where $m$ is the electron mass, $\hat{A}=\mathcal{A}(\hat{a}+\hat{a}^{\dagger})$ is the operator of the quantized cavity field and $\mathcal{A}$ is the mode amplitude. To simplify the problem, let us perform the successive unitary transformations. First, we apply the squeezing transformation $\hat{S}(\theta)=\exp[(\theta \hat{a}^2-\theta \hat{a}^{\dagger 2})/2]$, where $\tanh 2\theta = -1/(1+g^2/l_0^2)$, $l_0=\sqrt{\hbar/(m\Omega_0)}$ and $g=\hbar/e\mathcal{A}$. As a second step, we apply the quantized analogue of the Kramers-Henneberger transformation extensively used for the description of the interaction of atoms with intense classical electromagnetic fields~\cite{37,38}. This transformation corresponds to the transition to the noninertial frame of reference connected with the electron periodically driven by the electromagnetic field. For the case of quantized field, the transformation is given by the operator
\begin{align}
 \hat{\cal U}=\exp\left[-i\beta (-i\partial_x) i(\hat{a}^{\dagger}-\hat{a})\right], \label{Accel_transform}
\end{align}
where $\beta=\alpha (1+2\alpha^2)^{-3/4}$ and $\alpha=l_0/g$ is the dimensionless electron-photon coupling strength. Then the transformed Hamiltonian (\ref{H0}) reads
\begin{align}\label{H1}
    \hat{\cal H}_0^\prime=-\frac{\hbar^2}{2\widetilde{m}}\partial_x^2+V_0\delta(x-\sqrt{2}\xi\hat{q})+\frac{\hbar\Omega}{2}(\hat{\pi}^2+\hat{q}^2)
\end{align}
where $\widetilde{m}=m(1+2\alpha^2)$, $\hat{q} = i(\hat{a}^{\dagger}-\hat{a})/\sqrt{2}$, $\Omega=\Omega_0\sqrt{1+2\alpha^2}$ and $\hat{\pi}=(\hat{a}+\hat{a}^{\dagger})/\sqrt{2}$. It can be seen that the transformation (\ref{Accel_transform}) transfers the coupling between an electron and the electromagnetic field from the operator of electron kinetic energy to the scattering potential. We further normalize the Hamiltonian (\ref{H1}) to the energy $\hbar\Omega$ and normalize the electron coordinate $x$ to the length $\sqrt{\hbar/(\widetilde{m}\Omega)}$. The final dimensionless Hamiltonian (\ref{H1}) reads
\begin{align}
\hat{{\cal H}}=-\frac{1}{2}\partial_x^2 + \widetilde{V}\delta(x-\alpha \hat{q}) +\frac{1}{2}(\hat{q}^2+\hat{\pi}^2), \label{transformed_H}
\end{align}
where $\widetilde{V}=V_0(1+2\alpha^2)^{1/4}/(\hbar\Omega_0 l_0)$. It should be noted that the similar transformation has been recently used to analyze different light-matter coupling regimes for the electrons in periodic potentials~\cite{39}. The general solution of the Schr\"odinger equation with the Hamiltonian~\eqref{transformed_H} can be presented as a series
\begin{align}
\psi(x,q)=\sum_n \chi_n(x)\phi_n(q), ~\label{eq:fullWF}
\end{align}
where $\phi_n(q)$ are the eigenstates of the harmonic oscillator Hamiltonian. Substituting the  ansatz~\eqref{eq:fullWF} to the Schr\"odinger equation $\hat{\cal H}\psi=E\psi$, one can obtain a series of the one-dimensional differential equations on the amplitudes $\chi_n$.

Assuming the cavity to be initially in its ground state, the total electron-field wave function can be written as
\begin{align}\label {wf}
\psi(x,q) = e^{ipx}\phi_0(q) + \psi_{sc}(x,q),
\end{align}
where the first term corresponds to the incoming electron wave, the second one describes the scattered wave, $p=\sqrt{2E-1}$ is the electron momentum, and $E$ is the total energy. Taking into account the short range of the scattering potential, each electron basis function $\chi_n(x)$ can be approximated at $|x|\rightarrow \infty$ by the plane wave $\chi_n\sim e^{ip_n x}$ with the momentum $p_n=\sqrt{2(E-n)-1}$.  Therefore, there will be a finite number of partial amplitudes corresponding to $n<n_{max}=E-1/2$. In what follows, we will restrict the analysis by the energy $E<3/2$ so that there are not more than two partial amplitudes corresponding to $n=0$ and $n=1$. Thus, the asymptotic form of the scattered wave can be written as $\psi_{sc}=t_{00}e^{ipx}\phi_0(q) +t_{01}e^{ip_1x}\phi_1(q)$ and $\psi_{sc}=r_{00}e^{-ipx}\phi_0(q) +r_{01}e^{-ip_1x}\phi_1(q)$ for $x\to\pm\infty$, respectively. As a result, we arrive at the system of equations for the amplitudes $\chi_n$,
\begin{align}
\frac{-\partial^2_x\chi_n}{2} + \frac{\widetilde{V}}{\alpha}\sum_m \phi_m\left(\frac{x}{\alpha}\right)\phi_n\left(\frac{x}{\alpha}\right)\chi_m = \left(\frac{p^2}{2}-n\right)\chi_n. \label{sys_full}
\end{align}

\begin{figure}
    \centering
    \includegraphics[width=1.0\columnwidth]{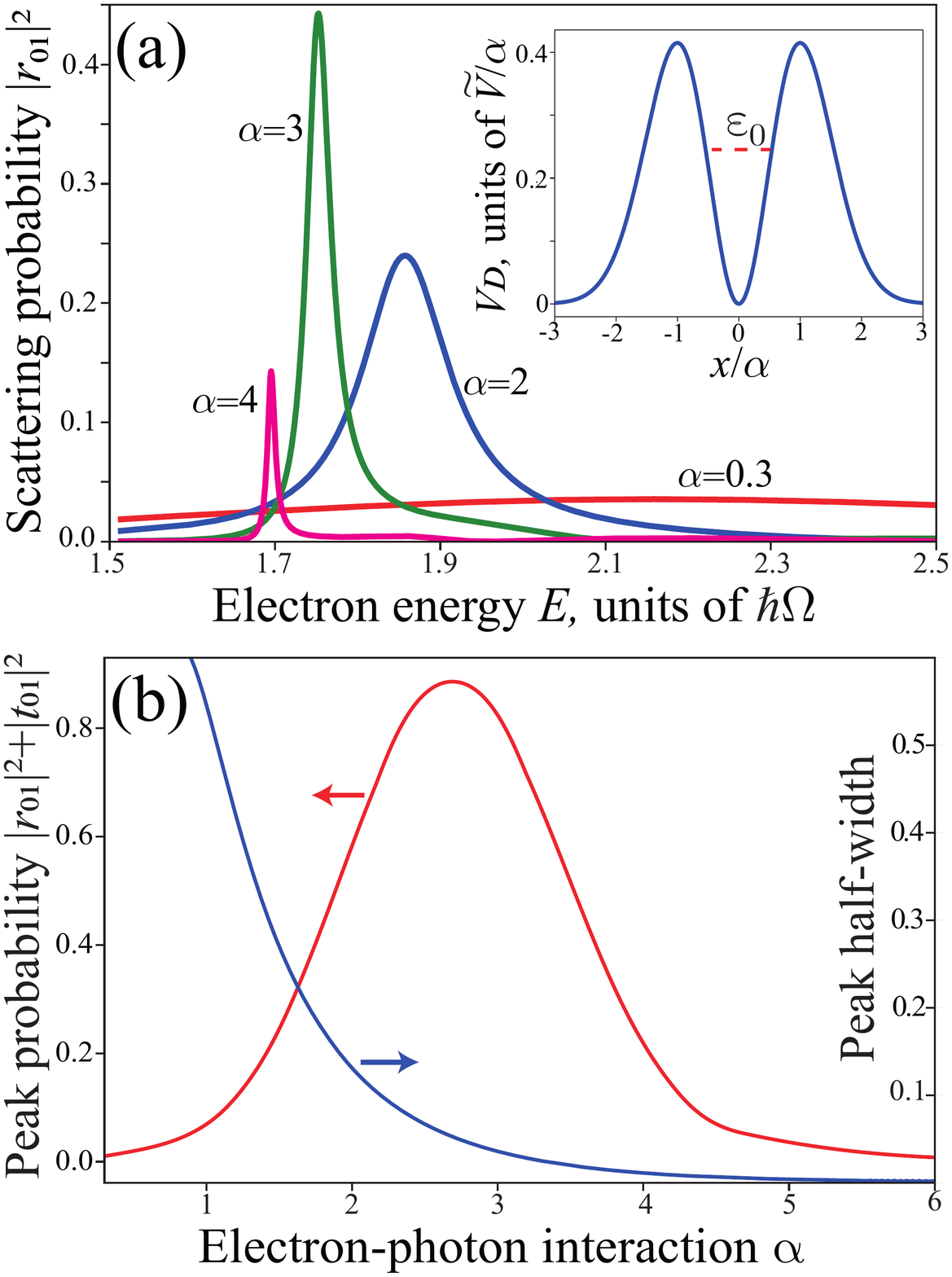}
    \caption{(a) The reflection scattering probability $|r_{01}|^2$ as a function of electron energy $E$ for the different electron-photon coupling strengths $\alpha$. The inset shows profile of the dressed scattering potential $V_{D}(x)$ containing the quasi-stationary polariton state with the energy $\varepsilon_{0}$ (the dashed line); (b) Dependencies of the peak value of the inelastic scattering probability and the peak half-width on the electron-photon coupling strength $\alpha$.
    }
    \label{fig_scattering}
\end{figure}

While the system of equations~(\ref{sys_full}) can be solved numerically for the arbitrary electron-photon interaction strength $\alpha$, let us obtain analytical approximations for the two limiting cases of $\alpha \ll 1$ and $\alpha \gg 1$. For the case of $\alpha \ll 1$, one can find that $\lim_{\alpha\rightarrow 0} \phi_0^2(x/\alpha)/\alpha= \delta(x)$ and $\lim_{\alpha\rightarrow 0}\phi_n(x/\alpha)\phi_{m}(x/\alpha)\sim \alpha^{|n-m|}$. As a consequence, the scattering-induced coupling of different eigenmodes of the cavity field is weak and can be treated perturbatively. Since the unperturbed (the zeroth-order of $\alpha$) amplitude is $\chi_0 = e^{ipx}-e^{ip|x|}(1-ip/\widetilde{V})^{-1}$, the amplitude $\chi_1$ in the principal order of $\alpha$ can be found as a solution of the equation
\begin{align}
\left[\frac{{p}_1^2}{2}+\frac{\partial_x^2}{2}-\frac{\widetilde{V}}{\alpha}\phi_1^2\left({x}/{\alpha}\right)\right]\chi_1=\frac{\widetilde{V}\phi_1\left({x}/{\alpha}\right)\phi_0\left({x}/{\alpha}\right)}{\alpha}\chi_0. \label{eq:main}
\end{align}
In the opposite limit of $\alpha\gg 1$, the coupling of different field eigenmodes is also weak due to the smallness of the ratio $\widetilde{V}/\alpha$. In the principal order of $1/\alpha$, the amplitude $\chi_1$ again satisfies Eq.~\eqref{eq:main}, where the amplitude $\chi_0$ can be approximated by a plane wave $e^{ipx}$ because $p^2/2\sim 1\gg \widetilde{V}/\alpha$.
It follows from the aforesaid that the scattering amplitude $\chi_1$ in the limits of both large and small $\alpha$ can be found by inverting the operator in the left-hand side of Eq.~\eqref{eq:main},
\begin{align}
    \chi_1(x)=\int dx' G_{1}(x,x')\frac{\tilde{V}}{\alpha}\phi_1\left(\frac{x'}{\alpha}\right)\phi_0\left(\frac{x'}{\alpha}\right)\chi_0(x'), \label{eq:sol}
\end{align}
where $G_{1}(x,x')$ is the Green's function for an electron in the potential $V_{D}(x)=(\widetilde{V}/\alpha) \phi_1^2(x/\alpha)$ which can be treated as a scattering potential renormalized by vacuum fluctuations of the cavity field (the dressed potential). Since the dressed potential $V_{D}(x)$ acquires the two-barrier structure (see the inset in Fig.~2a), it supports the polariton states confined between these two potential barriers at the specific energies $\varepsilon_0$. These states are quasi-stationary with the decay rate (the inverse lifetime of the state) $\gamma_0$ since they can decay due to the tunneling through the barriers.

While the exact values of the energies of the quasi-stationary states and the corresponding decay rates cannot be found analytically, certain approximations for the ground state can be made. Let us assume that the potential barriers confining the quasi-stationary states are sufficiently high. Then the potential $\phi_1^2$ can be approximated by a parabola. In this case, the ground state energy can be written as $\varepsilon_0\approx (\widetilde{V}/\alpha)(\sqrt{\pi}\alpha\widetilde{V})^{-1/2}$,  the barrier height reads $\varepsilon_{b}\approx\widetilde{V}/[\alpha \exp(1)]$, and the condition of the strong localization of the quasi-stationary state is $\alpha\widetilde{V} \gg 1$. Applying the well-known tunneling theory, the rate of tunneling decay reads $\gamma_0\approx \varepsilon_0\exp[-{\pi}\alpha(\varepsilon_b-\varepsilon_0)/4]$.
Existence of the quasi-stationary states leads to the resonant dependence of the scattering amplitude $\chi_1$ on the energy of the incoming electron. If the product $\alpha\tilde{V}$ decreases, the resonant state energy $\varepsilon_0$ approaches the barrier height $\varepsilon_b$, whereas the decay rate $\gamma_0$ approaches the state energy $\varepsilon_0$. As a result, the decreasing of the product $\alpha\widetilde{V}$ leads to vanishing the resonance. Correspondingly, the condition for the well pronounced resonance can be approximated by the inequality $\alpha \widetilde{V}>10$.

{\it Results and discussion.} The probability amplitudes for the electron scattering accompanied by emission of a cavity photon are determined by the inelastic reflection and transmission coefficients, $r_{01} = {e^{ip_1x}}\chi_1|_{x\rightarrow-\infty}$ and $t_{01}={e^{-ip_1x}}\chi_1|_{x\rightarrow\infty}$, respectively. The spectra of the scattering probability $|r_{01}|^2$ for the different values of $\alpha$, which are obtained from the exact numerical solution of the problem, are plotted in Fig.~2a. It is seen in the plots that both the resonant energy and the resonant peak half-width decrease with increasing $\alpha$. Qualitatively, this follows directly from the $\alpha$-dependence of the dressed barrier $V_{D}(x)$: Width of the dressed potential $V_D(x)$ increases with increasing $\alpha$, what decreases the resonant polariton energy $\varepsilon_{0}$ and increases the polariton lifetime. On the other hand, the right-hand side of Eq.~\eqref{eq:sol} at $\alpha \gg 1$ decreases with increasing $\alpha$ as $~e^{-p^2\alpha^2}$ since the amplitude $\chi_0(x)\sim e^{ipx}$ rapidly oscillates as compared to the increasingly smooth amplitude product $\phi_1(x/\alpha)\phi_0(x/\alpha)$. These opposing trends lead to the non-monotonic dependence of the peak probability of non-elastic scattering on the electron-photon interaction (see Fig.~2b). It is seen that the maximal value of the peak probability is achieved at $\alpha\approx 2.7$ that corresponds to the ultrastrong light-matter coupling regime~\cite{19}. Importantly, the maximal value of non-elastic reflection can reach the values of almost $0.5$ and thus the proposed geometry may be also interesting in the context of maximizing of quantum scattering by localized impurities, the topic which has been actively developing in recent years~\cite{39_1, 39_2}.

To estimate the feasibility of the proposed effect, let us consider a terahertz cavity with the frequency $\hbar\Omega_0=1$~meV, which corresponds to the length scale of $l_0=\sqrt{\hbar/(m\Omega_0)}\approx 8$~nm.  The electron-photon interaction {$\alpha$} can be increased by using the ultra-subwavelength cavities characterized by the effective cavity volume $\mathcal{V}=(\xi \lambda/2 )^3$, where $\lambda=2\pi c/\Omega_0$ is the resonant wavelength and $\xi$ is the dimensionless coefficient. Particularly, the condition $\alpha=1$ leads to the expression $\xi=(\hbar\Omega_0\alpha_0/(m\pi^3c^2))^{1/3}\approx 10^{-4}$, where $\alpha_0\approx1/137$ is the fine structure constant and the corresponding cavity lateral size is around $50$~nm. These values match to the recently obtained limit for the nanoconcentration of terahertz radiation in plasmonic cavities~\cite{40}. The required electron velocities in such cavities are of the order $10^4$~m/s, which is within the non-relativistic approximation and can be achieved with the compact free electron sources. For the case of near infrared cavities with the resonant frequency $\hbar\Omega_0\sim 1$~eV, the same analysis results in the required lateral cavity size of just several nanometers. Such extremely subwavelength cavities have been recently realized experimentally with the gap plasmonic modes~\cite{17} and are in principle feasible while still extremely challenging to fabricate.

It is also instructive to consider the spectral and temporal characteristics of the radiation emitted by a scattered electron. Immediately after the scattering, the system is in the linear superposition of the elastically and inelastically scattered waves, $\psi_E(x,q)= t_{00}e^{ipx}\phi_0(q)+t_{01}e^{i{p}_1x}\phi_1(q)$, where the second term corresponds to the excited polariton state and the total electron-field energy is $E$. To describe the dynamics of relaxation of this excited state, we adopt the approach of Ref.~\onlinecite{41}. Assuming a weak coupling to an external photonic reservoir and applying the Markov approximation, the output field operator $\hat{b}_{out}(t)$ can be represented as
$\hat{b}_{out}(t)=-\gamma \hat{X}^-(t)$,
where $\gamma$ is the cavity decay rate, and the operator $\hat{X}^-(0)=\sum_{E'<E} \langle \psi_{E'} | \hat{a}+\hat{a}^{\dagger}|\psi_E\rangle|\psi_{E'}\rangle\langle \psi_E|$ describes the relaxation between the eigenstates of the closed system. The final state $\psi_{E'}$ can be decomposed as  $|\psi_{E'}\rangle=\sum_{n}e^{ip_n(E')}\phi_n(q)$. The operator $\hat{X}^-(0)$ can be computed explicitly, yielding $\hat{X}^-(0) = (1+2\alpha^2)^{-1/4}\delta_{E',E-\hbar\Omega}$ and, thus, we can consider only two states. As a consequence, the master equation for the density matrix can be solved accurately. This allows to obtain the explicit expressions for the first order correlation function for the radiation $S_{t,\omega}$ defining the emission spectrum,
\begin{align}\label{S}
    &S_1(t,\omega)=\int d\tau e^{-i\omega \tau} \langle \hat{b}_{out}^{\dagger}(t+\tau)\hat{b}_{out}(t)\rangle=\nonumber \\&=e^{-\tilde{\gamma}t}(|t_{01}|^2+|r_{01}|^2)\frac{\tilde{\gamma}^2(1+2\alpha^2)^{-1/2}}{(\Omega-\omega)^2+\tilde{\gamma}^2/4},
\end{align}
where $\tilde{\gamma}=\gamma (1+2\alpha^2)^{-1/4}$.
\begin{figure}[!h]
    \centering
    \includegraphics[width=1.0\columnwidth]{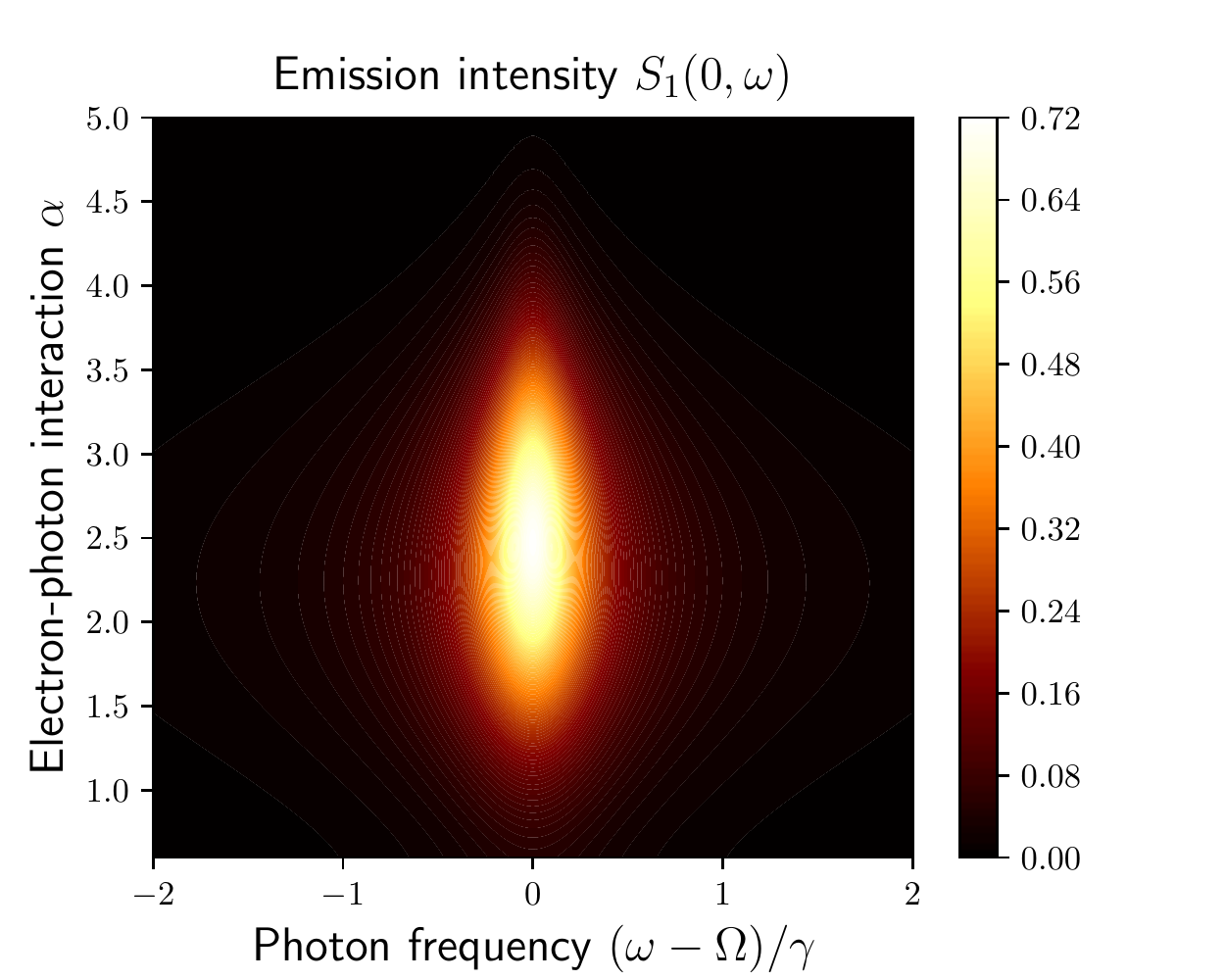}
    \caption{The intensity spectra of emitted radiation as a function of the electron-photon coupling strength $\alpha$.
    }\label{Fig:spectra}
\end{figure}
The map of the spectral intensity of the emission vs the coupling parameter $\alpha$ is plotted in Fig.~3. It follows from Eq.~(\ref{S}) that the emission peak width is suppressed by a factor $(1+2\alpha^2)^{1/4}$ which is approximately $2$ for the value of $\alpha\approx 2.7$ corresponding to the maximum of the inelastic transmission $t_{01}$.

It should be noted that the large values of electron-photon interaction crucial for the observation of the discussed effect can be achieved not only by the cavity design but also by introducing multiple electrons to the cavity. However, this would induce effective correlations between the electrons since all the electrons are coupled to the single cavity mode. At present, the interplay of the induced electron-electron correlations, the scattering from a static potential and the ultrastrong coupling to the cavity field remains so far open questions.

Recently, the interaction of free electrons with cavity vacuum fluctuations was studied experimentally~\cite{Copers1,Copers2}. In the experiments, a beam of free electrons, passing in the vicinity of a high-quality ring resonator, experienced a non-elastic scattering by the inhomogeneity of the cavity vacuum evanescent field. As a result of the scattering,  the generation of the antibunched cavity photons appears~\cite{Copers2}. It should be stressed that the observed effect is not resonant in the electron beam energy. On the contrary, we have shown that the formation of the resonant states via the dressing of the repulsive short-range potential leads to the resonant enhancement of the non-elastic scattering process when the electron energy matches the energy of the resonant state.

Finally, let us discuss the generalization of the considered 1D scattering problem to the 2D and 3D cases. In the case of a 3D repulsive potential, it requires three different cavity modes with non co-planar polarizations to produce the resonant electron state confined by the potential. Moreover, the vectorial nature of the cavity field precludes the formation of the spherically symmetric dressed potential. At the same time, one can use the  scattering potential elongated in one dimension, which leads to a cylindrically symmetric double barrier structure under the cavity dressing. To achieve this, a chiral cavity can be used, where the modes characterized by different circular polarizations have different frequencies. Lastly, if the potential is elongated along two dimensions (a wall-like potential), the analysis presented in the article can be applied directly.


\textit{Conclusion.} We have considered the scattering of a free electron on a short-range repulsive potential in the regime of ultrastrong light-matter coupling to a single-mode optical cavity. It is demonstrated that the dynamical stabilization of the electron at the potential by vacuum fluctuations of the electromagnetic field appears. Namely, the strong electron-photon interaction leads to the renormalization of the repulsive potential and its transformation to a double barrier potential hosting resonant scattering states. They result, particularly, in the resonant non-elastic scattering of the electron with emission of a cavity photon. Since this process can be very efficient for large values of electron-photon coupling, the present research opens the new avenue in the cavity QED engineering, establishing a link between the research on the interaction of free electrons with photonic nanostructures and the cavity QED.

\textit{Acknowledgements.} The reported study was funded by the Russian Science Foundation (Project No. 20-12-00001).  I.V.T. acknowledges funding by the Spanish Ministerio de Ciencia e Innovacion  (MICINN) (Project PID2020-112811GB-I00).

\end{document}